\documentstyle[12pt,epsf]{article}
\epsfverbosetrue
\def\PR #1 #2 #3 {Phys.~Rev.~{\bf #1}, #2 (#3)}
\def\PRL #1 #2 #3 {Phys.~Rev.~Lett.~{\bf #1}, #2 (#3)}
\def\PRD #1 #2 #3 {Phys.~Rev.~D~{\bf #1}, #2 (#3)}
\def\PLB #1 #2 #3 {Phys.~Lett.~{\bf B#1}, #2 (#3)}
\def\NPB #1 #2 #3 {Nucl.~Phys.~{\bf B#1}, #2 (#3)}
\def\RMP #1 #2 #3 {Rev.~Mod.~Phys.~{\bf #1}, #2 (#3)}
\def\ZPC #1 #2 #3 {Z.~Phys.~C~{\bf #1}, #2 (#3)}

\begin{document}
\begin{titlepage}

\rightline{hep-ph/9811492}
\rightline{DOE-ER-40757-121}
\rightline{UTEXAS-HEP-98-23}
\rightline{ILL-(TH)-98-5}
\rightline{ANL-HEP-PR-98-138} 
\rightline{EFI-98-57} 
\medskip
\rightline{November 1998}
\bigskip

\begin{center} {\Large \bf Higgs-Boson Production in Association with \\ 
\medskip Bottom Quarks at Next-to-Leading Order} \\
\bigskip\bigskip\bigskip\bigskip
{\large{\bf D.~Dicus}} \\ 
\medskip 
Center for Particle Physics \\ Department of Physics \\
University of Texas \\ Austin, TX\ \ 78712 \\
\bigskip\bigskip 
{\large{\bf T.~Stelzer}} \\ 
\medskip 
Department of Physics \\
University of Illinois \\ 1110 West Green Street \\  Urbana, IL\ \ 61801 \\
\bigskip\bigskip 
{\large{\bf Z.~Sullivan}} \\
\medskip
High Energy Physics Division \\
Argonne National Laboratory \\
Argonne, IL\ \ 60439 \\
\bigskip\bigskip 
{\large{\bf S.~Willenbrock}} \\ 
\medskip 
Enrico Fermi Institute and Department of Physics \\
University of Chicago \\ 5640 S.~Ellis Avenue \\
Chicago, IL\ \ 60637 \\
\smallskip
and\\
\smallskip
Department of Physics \\
University of Illinois \\ 1110 West Green Street \\  Urbana, IL\ \ 61801 \\
\bigskip 
\end{center} 
\bigskip\bigskip\bigskip

\begin{abstract}
We argue that the leading-order subprocess for Higgs-boson production in 
association with bottom quarks is $b\bar b\to H$.  This process is an 
important source of Higgs bosons with enhanced Yukawa coupling to 
bottom quarks.  We calculate the corrections to this cross section at 
next-to-leading-order in $1/\ln (m_H/m_b)$ and $\alpha_s$  
and at next-to-next-to-leading order in $1/\ln (m_H/m_b)$. 
\end{abstract}

\end{titlepage}

\newpage

\section{Introduction}

\indent\indent
The standard-model Higgs boson has a very weak Yukawa coupling to bottom 
quarks, proportional to $m_b/v$, where $v \approx 246$ GeV is the 
vacuum-expectation 
value of the Higgs field.  Therefore, the cross section for the production 
of the standard-model Higgs boson in association with bottom quarks is 
relatively small at the Fermilab Tevatron 
\cite{SMW} and the CERN Large Hadron Collider (LHC) \cite{DW}, in comparison
with other Higgs-boson production cross sections.  However, 
if the bottom-quark Yukawa coupling is enhanced, this production mechanism 
can be a significant source of Higgs bosons \cite{DW}.
Such an enhancement occurs, for example, in a two-Higgs-doublet model
for large values of $\tan\beta \equiv v_2/v_1$, where $v_{1,2}$ are the 
vacuum-expectation values of the two Higgs fields.  A value as large as 
$\tan\beta\approx m_t/m_b$ is obtained in the simplest version of an 
SO(10) grand-unified theory.  The Higgs boson may
be detected via its decay to $\tau^+\tau^-$ \cite{KZ,DGR,CMW} 
or $b\bar b$ \cite{CMW,DGV,DHTY,CDR} at the LHC and the Tevatron,
and $\mu^+\mu^-$ \cite{KS} at the LHC.

In this paper we calculate the cross section for Higgs-boson
production in association with bottom quarks at next-to-leading order.
We argue that the leading-order subprocess is $b\bar b \to H$, where
the initial $b$-quark distribution function is calculated
\cite{OT,BHS}.  We show that the subprocess $gb \to Hb$ is a
correction to the leading-order subprocess of order $1/\ln(m_H/m_b)$,
and that the subprocess $gg \to b\bar bH$ is a correction of order
$1/\ln^2(m_H/m_b)$. We calculate both of these corrections, and
confirm the calculation performed by two of us ten years ago
\cite{DW}.  Our new contribution to this part of the calculation is a
proper understanding of the relative order of the different
subprocesses.

Once we properly identify $b\bar b \to H$ as the leading-order subprocess, 
it is straightforward to calculate the order $\alpha_s$ correction 
from the emission of virtual and real gluons.  This calculation is performed
here for the first time.  We thus obtain the cross section for
Higgs-boson production in association with bottom quarks at next-to-leading 
order in both $\alpha_s$ and $1/\ln(m_H/m_b)$, as well as at 
next-to-next-to-leading order in $1/\ln(m_H/m_b)$. Our calculation is valid
for both scalar and pseudoscalar Higgs bosons. 

Our calculation corresponds to the inclusive cross section for Higgs-boson
production in association with bottom quarks, integrated over the momenta of
the $b$ quarks.  It is appropriate to use our results to normalize the 
inclusive cross section from a shower Monte Carlo program, such as PYTHIA
or HERWIG, which uses $b\bar b\to H$ as the hard-scattering 
subprocess.\footnote{PYTHIA and HERWIG use backwards evolution of the 
initial-state $b$ distribution functions to give the initial $g\to b\bar b$
splitting.}

The remainder of this paper is organized as follows.  In
Sec. \ref{sec_2} we explain how to properly count the order of the
various contributions to Higgs-boson production in association with
heavy quarks.  In Sec. \ref{sec_3} we perform the calculation of the
$1/\ln(m_H/m_Q)$ and $1/\ln^2(m_H/m_Q)$ corrections.  In
Sec. \ref{sec_4} we calculate the $\alpha_s$ correction.  In
Sec. \ref{sec_5} we present numerical results at the Tevatron and the
LHC.
  
\section{Counting orders}\label{sec_2}

\subsection{$1/\ln (m_H/m_Q)$ correction}\label{sec_2a}

\indent\indent In this section we explain how to count the order of the
various contributions to Higgs-boson production in association with heavy 
quarks.  This counting is a generalization, to the case with two heavy
quarks in the initial state, of the counting developed in Ref.~\cite{SSW}
for a subprocess with one heavy quark in the initial state ($qb \to qt$).
The underlying concepts for the organization of the calculation were 
developed in Refs.~\cite{OT,BHS}.

The actual physical subprocess for Higgs-boson production in association
with heavy quarks is $gg \to Q\overline QH$, shown in Fig.~1. 
Imagine that the heavy quark is very light compared 
with the Higgs-boson.  When the initial gluon splits into a 
nearly-collinear $Q\overline Q$ pair, the amplitude is enhanced by the 
propagator 
of the internal heavy quark, which is nearly on-shell.  Integrating over the 
phase space of the external heavy quark yields a factor of $\ln (m_H/m_Q)$, so 
the splitting of a gluon into a $Q\overline Q$ pair is intrinsically of order
$\alpha_s\ln (m_H/m_Q)$ (for $m_H \gg m_Q$).  Such a splitting occurs twice
in each of the first two diagrams of Fig.~1, once in each of the next four
diagrams,\footnote{In the middle 
four diagrams, one gluon splits into a $Q\overline Q$ pair, the other into 
$Q\overline QH$. Only the former gives rise to a factor of $\ln (m_H/m_Q)$.}
and not at all in the final two diagrams.

\begin{figure}[tb]
\begin{center}
\epsfxsize= 2.25in  %\epsfxsize= 2.2in  %actual
\leavevmode
\epsfbox{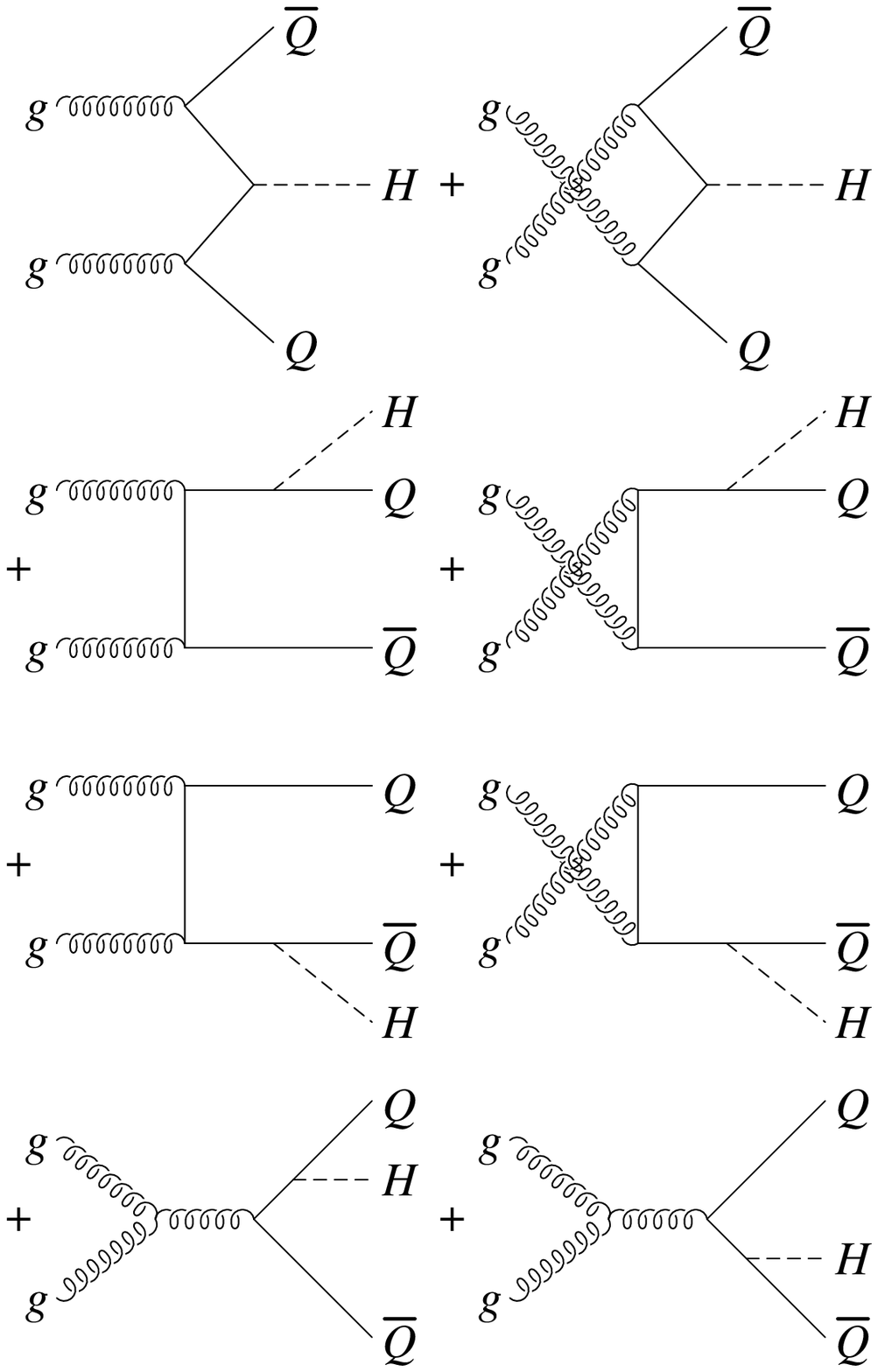}
\caption{Feynman diagrams for $gg \to Q\overline QH$.}
\end{center}
\end{figure}

Another power of this logarithm appears at every order in 
perturbation theory, via the emission of a collinear gluon from the 
nearly-on-shell quark propagator.  Thus the expansion parameter is 
$\alpha_s\ln (m_H/m_Q)$, and since the logarithm is large, the convergence
of the expansion is degraded.

The convergence of the expansion is improved by summing these
collinear logarithms to all orders in perturbation theory
\cite{OT,BHS}.  This is achieved by introducing a
(theoretically-defined) heavy-quark distribution function, $Q(x,\mu)$,
and using the Dokshitzer-Gribov-Lipatov-Altarelli-Parisi (DGLAP)
equations to sum the collinear logarithms.  The heavy-quark
distribution function is intrinsically of order $\alpha_s\ln
(m_H/m_Q)$ since it arises from the splitting of a gluon into a
nearly-collinear $Q\overline Q$ pair \cite{SSW}.

Once a heavy-quark distribution function is introduced, it changes the
way perturbation theory is ordered.  The leading-order subprocess is
$Q\overline Q \to H$, as shown in Fig.~2(a).  It is intrinsically of
order $\alpha_s^2\ln^2 (m_H/m_Q)$, since each heavy-quark distribution
function is of order $\alpha_s\ln (m_H/m_Q)$.  (There is also a factor
of the heavy-quark Yukawa coupling, which we suppress throughout this
discussion.)

\begin{figure}[tb]
\begin{center}
\epsfxsize= 3.0in
\leavevmode
\epsfbox{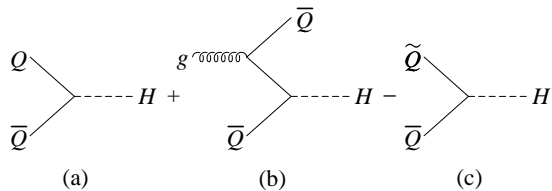}
\caption{Feynman diagrams for (a) the leading-order subprocess 
$Q\overline Q \to H$; (b) $g\overline Q \to H\overline Q$
(there is also an $s$-channel diagram, not shown); and 
(c) $\widetilde Q\overline Q \to H$, where the heavy-quark distribution 
function 
$\widetilde Q$ is given by the perturbative solution to the DGLAP equations.
Figs.~(b) and (c) together constitute the $1/\ln (m_H/m_Q)$ correction to 
the leading-order subprocess in (a).}
\end{center}
\end{figure}

Consider next the subprocess $g\overline Q \to H\overline Q$ 
(and related subprocesses), shown in 
Fig.~2(b).  This subprocess gives rise to a factor of $\ln (m_H/m_Q)$
from the region where the gluon splits into a nearly-collinear $Q\overline Q$ 
pair.
However, this logarithm has been summed into the heavy-quark distribution
function in Fig.~2(a), so it must be removed.  This is achieved by 
subtracting the diagram in Fig.~2(c), which corresponds to the subprocess
$Q\overline Q \to H$, but with the heavy-quark distribution function given by 
the perturbative solution to the DGLAP equation for a gluon splitting to a
$Q\overline Q$ pair,
\begin{equation}
\widetilde Q(x,\mu) = 
\frac{\alpha_s(\mu)}{2\pi}\ln \left(\frac{\mu^2}{m_Q^2}\right) 
\int_x^1 \frac{dy}{y}
P_{qg}\left(\frac{x}{y}\right)g(y,\mu) \;,
\label{Q}
\end{equation}
where $g$ is the gluon distribution function, and the DGLAP splitting 
function is given by
\begin{equation}
P_{qg}(z)=\frac{1}{2}[z^2+(1-z)^2]\;.
\end{equation}
After the cancellation of the logarithm, the sum of the subprocesses in
Figs.~2(b) and (c) is of order $\alpha_s$ times the order of the other 
heavy-quark distribution function, {\it i.e.}, of order 
$\alpha_s^2\ln (m_H/m_Q)$.  This is down by one power of $\ln (m_H/m_Q)$
with respect to the leading-order subprocess, Fig.~2(a), so it is a 
correction of order $1/\ln (m_H/m_Q)$.

Now consider the subprocess $gg\to Q\overline QH$, shown in Fig.~3(a).  This 
subprocess gives rise to a factor of $\ln (m_H/m_Q)$ when either gluon 
splits into
a nearly-collinear $Q\overline Q$ pair.  Since these collinear logarithms 
have been
summed into the heavy-quark distribution functions, they must be subtracted.
This is shown in Figs.~3(b)--3(d): each of the two collinear regions 
must be subtracted, but we must also add back the double-collinear region
[Fig.~3(d)], which is subtracted twice [Figs.~3(b) and (c)].  
Once the logarithms have been cancelled, the sum of the subprocesses in Fig.~3
is of order $\alpha_s^2$.  This is down by two powers of $\ln (m_H/m_Q)$
with respect to the leading-order subprocess, Fig.~2(a), so it is a 
correction of order $1/\ln^2 (m_H/m_Q)$.

\begin{figure}[tb]
\begin{center}
\epsfxsize= 2.25in  %\epsfxsize= 2.2in  %actual
\leavevmode
\epsfbox{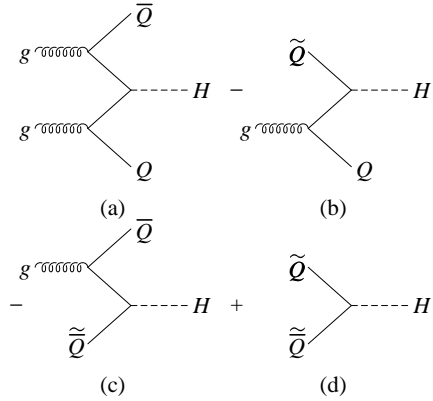}
\caption{Feynman diagrams for (a) $gg\to Q\overline QH$ (the complete set of 
diagrams is shown in Fig.~1); 
(b),(c) $\widetilde Qg \to QH$ and $g\widetilde{\overline Q} \to H\overline Q$
(there are also $s$-channel diagrams, not shown); and 
(d) $\widetilde Q\widetilde{\overline Q} \to H$, 
where the heavy-quark distribution function 
$\widetilde Q$ is given by the perturbative solution to the DGLAP equations.
These diagrams together constitute the $1/\ln^2 (m_H/m_Q)$ correction to 
the leading-order subprocess $Q\overline Q \to H$.}
\end{center}
\end{figure}

Thus we see that Higgs-boson production in association with heavy quarks
contains terms of relative order unity, $1/\ln (m_H/m_Q)$, and 
$1/\ln^2 (m_H/m_Q)$, depending on whether the initial state contains two,
one, or zero heavy quarks, respectively.  These are the only powers of
$1/\ln (m_H/m_Q)$ that appear, to all orders in perturbation theory \cite{SSW}.

\subsection{$\alpha_s$ correction}\label{sec_2b}

\indent\indent Now consider the correction to the leading-order subprocess,
$Q\overline Q\to H$ [Fig.~2(a)], 
from virtual- and real-gluon emission, shown in Fig.~4.
Since these diagrams contain two heavy quarks in the initial state, they
are of order $\alpha_s^3 \ln^2 (m_H/m_Q)$, {\it i.e.}, down by one power
of $\alpha_s$ from the leading-order subprocess.  

\begin{figure}[tb]
\begin{center}
\epsfxsize= 3.0in
\leavevmode
\epsfbox{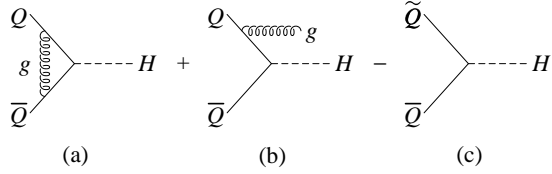}
\caption{Feynman diagrams for (a) the virtual-gluon correction to 
$Q\overline Q \to H$; (b) $Q\overline Q \to Hg$; and (c) $\widetilde Q
\overline Q \to H$, 
where the heavy-quark distribution function 
$\widetilde Q$ is given by the perturbative solution to the DGLAP equations.
These diagrams together constitute the $\alpha_s$ correction to the 
leading-order subprocess $Q\overline Q \to H$.}
\end{center}
\end{figure}

There is a factor of $\ln (m_H/m_Q)$ associated with the emission of a
collinear gluon from a heavy quark [Fig.~4(b)], and this is handled in
a similar manner to the collinear logarithm associated with a gluon
splitting to a $Q\overline Q$ pair.  The collinear logarithm is
summed, to all orders in perturbation theory, into the heavy-quark
distribution function, and the collinear region is then explicitly
removed by subtracting the subprocess $Q\overline Q\to H$ [Fig.~4(c)],
with the heavy-quark distribution function given by the perturbative
solution to the DGLAP equation for a gluon radiated from a heavy
quark.  After the cancellation of the collinear logarithms, the sum of
Figs.~4(b) and (c) [as well as Fig.~4(a)] is a correction of order
$\alpha_s$ to the leading-order subprocess $Q\overline Q\to H$.

\section{The $1/\ln (m_H/m_Q)$ correction}\label{sec_3}

\indent\indent The leading-order hadronic cross section for Higgs-boson 
production in 
association with heavy quarks, the $1/\ln (m_H/m_Q)$ correction, and the
$1/\ln^2 (m_H/m_Q)$ correction are obtained from the equations
\begin{eqnarray}
\sigma_{LO} 
& = & Q \otimes \sigma^0_{Q\overline Q \to H} \otimes \overline Q
    + \overline Q \otimes \sigma^0_{\overline QQ \to H} \otimes Q 
\label{LO}
\\ \nonumber \\
\sigma_{1/\ln (m_H/m_Q)}
& = & g \otimes \sigma^1_{gQ \to HQ} \otimes Q 
    - \widetilde{\overline Q} \otimes \sigma^0_{\overline QQ \to H} \otimes Q 
\nonumber \\
& + & Q \otimes \sigma^1_{Qg \to QH} \otimes g
    - Q \otimes \sigma^0_{Q\overline Q \to H} \otimes \widetilde{\overline Q}
\nonumber \\
& + & g \otimes \sigma^1_{g\overline Q \to H\overline Q} \otimes \overline Q
    - \widetilde Q \otimes \sigma^0_{Q\overline Q \to H} \otimes \overline Q
\nonumber \\
& + & \overline Q \otimes \sigma^1_{\overline Qg \to \overline QH} \otimes g 
    - \overline Q \otimes \sigma^0_{\overline QQ \to H} \otimes \widetilde Q
\label{NLO}
\\ \nonumber \\
\sigma_{1/\ln^2 (m_H/m_Q)}
& = & g \otimes \sigma^2_{gg\to Q\overline QH} \otimes g 
\nonumber \\
& - & \widetilde Q \otimes \sigma^1_{Qg \to QH} \otimes g
    - g \otimes \sigma^1_{gQ \to HQ} \otimes \widetilde Q
\nonumber \\
& - & \widetilde{\overline Q} \otimes \sigma^1_{\overline Qg \to \overline QH}
\otimes g 
    - g \otimes \sigma^1_{g\overline Q \to H\overline Q}\otimes 
\widetilde{\overline Q}
\nonumber \\
& + & \widetilde Q \otimes \sigma^0_{Q\overline Q \to H} \otimes 
\widetilde{\overline Q}
    + \widetilde{\overline Q} \otimes \sigma^0_{\overline QQ \to H} \otimes 
\widetilde Q
\label{NNLO}
\end{eqnarray}
where $Q$ is the heavy-quark distribution function, $\widetilde Q$ is
the perturbative heavy-quark distribution function [Eq.~(\ref{Q})],
$g$ is the gluon distribution function, and they are convolved with
the various subprocess cross sections in the usual way.  (The parton
distribution function written before the subprocess cross section is
from hadron A, the one written after from hadron B, and the
superscripts on the subprocess cross sections denote the power of
$\alpha_s$.)  This formula implements the discussion in
Sec. \ref{sec_2} on the proper way to count the order of the
contributions to Higgs-boson production in association with heavy
quarks.\footnote{The next-to-next-to-leading-order formula,
Eq.~(\ref{NNLO}), not only subtracts the double-collinear region, as
described in Sec. \ref{sec_2a}, but also subtracts the
single-collinear regions.}  One can check that the sum of these
equations is equivalent to Eq.~(5) of Ref.~\cite{DW}, although the
proper way to count orders was not known at that time.

For the calculation of the $1/\ln (m_H/m_Q)$ correction (and also the
$\alpha_s$ correction), it is more convenient to regulate the
collinear divergence with dimensional regularization \cite{BHS} rather
than with a finite heavy-quark mass \cite{OT}.  The former is accurate
up to powers of $m_Q^2/m_H^2$, which is small in the region of
validity of our calculation, $m_Q \ll m_H$.  We perform the
calculation of the $1/\ln (m_H/m_Q)$ correction both ways,
analytically in the case of dimensional regularization and numerically
in the case of a finite heavy-quark mass, and find agreement.  The
calculation of the $1/\ln^2 (m_H/m_Q)$ correction is only done
numerically, using a finite heavy-quark mass.

We now describe the analytic calculation of the $1/\ln (m_H/m_Q)$
correction using dimensional regularization.  The calculation is
similar to the QCD correction to the Drell-Yan process from initial
gluons \cite{AEM},\footnote{For a pedagogical treatment, see
Ref.~\cite{W}.} but with the vector current replaced by a scalar
current.

The spin- and color-averaged cross section for the leading-order
subprocess $Q\overline Q\to H$ is
\begin{equation}
\sigma^0_{Q\overline Q \to H} = \frac{\pi}{6}
	\frac{\overline m^2(\mu)}{v^2}\mu^{2\epsilon}
	\frac{1}{m_H^2}\delta(1-z) \;,
\label{B}
\end{equation}
where $z\equiv m_H^2/\hat{s}$.  The calculation is performed in
$N=4-2\epsilon$ dimensions; $\mu$ is the `t Hooft mass, which is
introduced such that the renormalized Yukawa coupling is dimensionless
in $N$ dimensions.  We use the $\overline {\rm MS}$ scheme to subtract
ultraviolet (and collinear) divergences, so $\overline m(\mu)$ is the
heavy-quark $\overline {\rm MS}$ mass.

The spin- and color-averaged cross section for the subprocess $gQ \to HQ$ is
\begin{equation}
\sigma^1_{gQ \to HQ} = \frac{\alpha_s}{12}\frac{\overline m^2(\mu)}{v^2}
\mu^{4\epsilon}
\frac{1}{\hat{s}}
	\Biggl\{P_{qg}(z)
\Biggl[-\frac{1}{\epsilon}\frac{\Gamma(1-\epsilon)}{\Gamma(1-2\epsilon)}
+\ln\Biggl(\frac{m_H^2}{4\pi}
	\frac{(1-z)^2}{z}\Biggr)\Biggr]+\frac{1}{4}(1-z)(7z-3)\Biggr\} \;,
\label{I}
\end{equation}
where $z\equiv m_H^2/\hat{s}$.  The collinear divergence 
manifests itself as a pole at 
$\epsilon = 0$.  The `t Hooft mass accompanies both the renormalized strong 
coupling and Yukawa coupling, which are defined to be dimensionless in 
$N$ dimensions.

The perturbative heavy-quark distribution function that subtracts
the collinear region in dimensional regularization is
\begin{equation}
\widetilde Q(x,\mu) = - \frac{\alpha_s}{2\pi}\int_x^1 \frac{dy}{y} 
P_{qg}\left(\frac{x}{y}\right) g(y,\mu)
\left(\frac{1}{\epsilon}-\gamma+\ln 4\pi\right)\;.
\label{MSI}
\end{equation}
This is the analogue of Eq.~(\ref{Q}), in which 
the collinear divergence is regularized by a finite heavy-quark mass. 
Either distribution function can be used in Eq.~(\ref{NLO}), and both yield
the cross section in the $\overline {\rm MS}$ scheme.  Using dimensional 
regularization, the 
first line of Eq.~(\ref{NLO}) yields the cross section 
\begin{equation}
\bar\sigma_{gQ \to HQ}^1 = \frac{\alpha_s}{12}
\frac{\overline m^2(\mu)}{v^2}
\frac{1}{\hat{s}}
	\left[P_{qg}(z)\ln\left(\frac{m_H^2}{\mu^2}
	\frac{(1-z)^2}{z}\right)+\frac{1}{4}(1-z)(7z-3)\right]\;.
\end{equation}
This is just Eq.~(\ref{I}) with the term proportional to 
$(1/\epsilon - \gamma + \ln 4\pi)$ removed by renormalization, and the
limit $\epsilon \to 0$ taken.  The remaining terms in the $1/\ln (m_H/m_Q)$ 
correction [lines 2--4 of Eq.~(\ref{NLO})] yield the same expression.

The $1/\ln^2 (m_H/m_Q)$ correction, Eq.~(\ref{NNLO}), can also be
calculated analytically using dimensional regularization.  However, we
find it simpler to perform the calculation numerically, using a finite
heavy-quark mass.  The perturbative heavy-quark distribution function
that subtracts the collinear region is given by Eq.~(\ref{Q}), and
$m_Q$ is kept finite throughout the calculation.

\section{The $\alpha_s$ correction}\label{sec_4}

\indent\indent The calculation of the $\alpha_s$ correction to $Q\overline Q 
\to H$ is also similar
to the correction to the Drell-Yan process \cite{AEM,W}.  However, there is an 
additional feature: the ultraviolet renormalization of the Yukawa coupling
\cite{BL}.  
The electroweak coupling is not renormalized in the Drell-Yan process due to a
Ward identity which cancels the ultraviolet divergence.

The interference of the one-loop vertex correction in Fig.~4(a) with the 
tree diagram in Fig.~2(a) yields the spin- and color-averaged cross section
\begin{equation}
\sigma^1_{Q\overline Q \to H} = \frac{\pi}{6}
	\frac{\overline m^2(\mu)}{v^2}\mu^{2\epsilon}
	\frac{1}{\hat{s}}\delta(1-z)
	\left[1+C_F\frac{\alpha_s}{2\pi}\mu^{2\epsilon}
	\left(\frac{4\pi}{m_H^2}\right)^\epsilon
	\frac{\Gamma(1-\epsilon)}{\Gamma(1-2\epsilon)}
	\left(-\frac{2}{\epsilon^2}-\frac{3}{\epsilon}-2
	+\frac{2\pi^2}{3}\right)\right] \;,
\label{BV}
\end{equation}
($C_F \equiv 4/3$)
which includes both the leading-order and next-to-leading-order terms.  
The Yukawa coupling has been renormalized in the $\overline{\rm MS}$ 
scheme, which yields the counterterm \cite{BL}
\begin{equation}
{\cal L} = -\mu^\epsilon\frac{\overline m(\mu)}{v}
\left(1-\frac{\delta \overline m}{\overline m}\right) Q\overline Q H \;,
\end{equation}
where 
\begin{equation}
\frac{\delta \overline m}{\overline m} = C_F\frac{\alpha_s}{4\pi}3
\left(\frac{1}{\epsilon}-\gamma+\ln 4\pi\right)\;.  
\end{equation}
The cross section displays both a collinear ($1/\epsilon$) and an infrared
($1/\epsilon^2$) divergence.

The emission of a real gluon ($Q\overline Q \to Hg$) yields the spin- and 
color-averaged cross section
\begin{eqnarray}
\sigma^1_{Q\overline Q \to Hg} 
& = & C_F\frac{\alpha_s}{12}\mu^{2\epsilon}\frac{\overline m^2(\mu)}{v^2}
	\frac{1}{\hat{s}}
	\left(\frac{4\pi}{m_H^2}\right)^\epsilon
	\frac{\Gamma(1-\epsilon)}{\Gamma(1-2\epsilon)} \nonumber \\
	&\times&\left[\frac{2}{\epsilon^2}\delta(1-z)
	-\frac{2}{\epsilon}\frac{1+z^2}{(1-z)_+} 
	-2(1+z^2)\frac{\ln z}{1-z}+4(1+z^2)\left(\frac{\ln(1-z)}{1-z}\right)_+
         +2(1-z)\right] \;, \nonumber \\
\label{R}
\end{eqnarray}
where $z\equiv m_H^2/\hat{s}$, and the ``plus'' prescription is defined as
usual:
\begin{equation}
\int_0^1 dz\, [f(z)]_+ h(z) \equiv \int_0^1 dz\, f(z) [h(z)-h(1)]\;.
\end{equation}  
When combined with the cross section from virtual-gluon emission, 
Eq.~(\ref{BV}), the infrared divergences cancel.
The collinear divergence is subtracted by constructing the combination
\begin{eqnarray}
\sigma_{\alpha_s} 
& = & Q \otimes \sigma^1_{Q\overline Q \to H} \otimes \overline Q
    + \overline Q \otimes \sigma^1_{\overline QQ \to H} \otimes Q 
\nonumber \\
& + & Q \otimes \sigma^1_{Q\overline Q \to Hg} \otimes \overline Q 
    - \widetilde Q \otimes \sigma^0_{Q\overline Q \to H} \otimes \overline Q 
    - Q \otimes \sigma^0_{Q\overline Q \to H} \otimes \widetilde{\overline Q} 
\nonumber \\
& + & \overline Q \otimes \sigma^1_{\overline QQ \to Hg} \otimes Q 
    - \widetilde{\overline Q} \otimes \sigma^0_{\overline QQ \to H} \otimes Q 
    - \overline Q \otimes \sigma^0_{\overline QQ \to H} \otimes \widetilde Q 
\;,
\label{NLOA}
\end{eqnarray}
which is the analogue, for virtual- and real-gluon emission, of 
Eq.~(\ref{NLO}).  The perturbative heavy-quark distribution 
function in Eq.~(\ref{NLOA}) is given by 
\begin{equation}
\widetilde Q(x,\mu) = - \frac{\alpha_s}{2\pi}\int_x^1 \frac{dy}{y} 
P_{qq}\left(\frac{x}{y}\right) Q(y,\mu)
\left(\frac{1}{\epsilon}-\gamma+\ln 4\pi\right) \;,
\end{equation}
where
\begin{equation}
P_{qq}(z)=C_F\left(\frac{1+z^2}{(1-z)_+} +
\frac{3}{2}\delta(1-z)\right) \;,
\end{equation}
is the DGLAP splitting function for a quark radiating a gluon.  The sum
of virtual- and real-gluon emission, after the subtraction of the collinear
divergence, is
\begin{eqnarray}
\bar\sigma^1_{Q\overline Q \to H} + \bar\sigma^1_{Q\overline Q \to Hg} 
& = & \frac{\pi}{6}
	\frac{\overline m^2(\mu)}{v^2}\frac{1}{\hat{s}} \nonumber \\
       &\times&\Biggl\{ \delta(1-z) 
+\frac{\alpha_s}{\pi}P_{qq}(z)\ln\frac{m_H^2}{\mu^2}\nonumber \\
       &&+C_F\frac{\alpha_s}{2\pi}\Biggl[\Biggl(-2+\frac{2\pi^2}{3}
          -3\ln\frac{m_H^2}{\mu^2}\Biggr) \delta(1-z) \nonumber \\
       &&-2(1+z^2)\frac{\ln z}{1-z}+4(1+z^2)\Biggl
(\frac{\ln(1-z)}{1-z}\Biggr)_+
         +2(1-z)\Biggr]\Biggr\} \;,
\label{BVR}
\end{eqnarray}
which is the sum of Eqs.~(\ref{BV}) and (\ref{R}), with the term proportional
to $(1/\epsilon - \gamma + \ln 4\pi)$ removed by renormalization, and
the limit $\epsilon \to 0$ taken.

Since the derivation of Eq.~(\ref{BVR}) involves the removal of both 
collinear and ultraviolet divergences, there are actually two independent 
scales ($\mu$) present.  To make this explicit, consider the leading-order 
running of the heavy-quark mass and the strong coupling \cite{BL}:
\begin{equation}
\overline m(\mu) = \overline m(\mu_0)
\left(\frac{\alpha_s(\mu)}{\alpha_s(\mu_0)}\right)^{\frac{1}{\beta_0}}
\;,
\label{BRNG}
\end{equation}
\begin{equation}
\alpha_s(\mu) = \frac{\alpha_s(\mu_0)}
{1+\beta_0[\alpha_s(\mu_0)/\pi]\ln (\mu^2/\mu_0^2)} \;,
\end{equation}
where $\beta_0 = (11-2N_f/3)/4$
(see the Appendix for the next-to-leading-order equation).
The perturbative expansion of Eq.~(\ref{BRNG}) at next-to-leading order is
\begin{equation}
\overline m(\mu) = \overline m(\mu_0)\left[1-C_F\frac{\alpha_s}{4\pi}3
\ln\left(\frac{\mu^2}{\mu_0^2}\right)\right]\;.
\label{RNG}
\end{equation}
Using Eq.~(\ref{RNG}) to replace $\overline m(\mu)$ with $\overline 
m(\mu_{UV})$ in Eq.~(\ref{BVR}) yields our final expression for the 
correction from virtual- and real-gluon emission,
\begin{eqnarray}
\bar\sigma^1_{Q\overline Q \to H} + \bar\sigma^1_{Q\overline Q \to Hg} 
& = & \frac{\pi}{6}
	\frac{\overline m^2(\mu_{UV})}{v^2}\frac{1}{\hat{s}} \nonumber \\
       &\times&\Biggl\{ \delta(1-z) 
+\frac{\alpha_s}{\pi}P_{qq}(z)\ln\frac{m_H^2}{\mu^2}\nonumber \\
       &&+C_F\frac{\alpha_s}{2\pi}\Biggl[\Biggl(-2+\frac{2\pi^2}{3}
          -3\ln\frac{m_H^2}{\mu_{UV}^2}\Biggr) \delta(1-z) \nonumber \\
       &&-2(1+z^2)\frac{\ln z}{1-z}+4(1+z^2)\left(\frac{\ln(1-z)}{1-z}\right)_+
         +2(1-z)\Biggr]\Biggr\} \;,
\end{eqnarray}
where we now distinguish between the renormalization scale ($\mu_{UV}$), 
associated with the 
running coupling, and the factorization scale ($\mu$), associated with 
the parton distribution functions.

\section{Results and Conclusions}\label{sec_5}

\indent\indent Our numerical results for Higgs-boson production in
association with bottom quarks at the Tevatron ($\sqrt S = 1.8$ and 2
TeV $p\bar p$) and the LHC ($\sqrt S = 14$ TeV $pp$) are presented in Table
\ref{tab_one}.\footnote{The contribution from $q\bar q \to b\bar b H$
is negligible at both machines.}  All cross sections are evaluated
with the CTEQ4M parton distribution functions \cite{CTEQ}.  The
factorization ($\mu$) and renormalization ($\mu_{UV}$) scales are both
set equal to $m_H$.  The running $b$ mass is evolved from an initial
value of $\overline m_b(M_b) = 4.25 \pm 0.15$ GeV
\cite{PDG}\footnote{$M_b$ is the pole mass, which equals 4.64 GeV at
one loop.  This is the appropriate value to use in $\overline
m_b(M_b)$, since the $\overline {\rm MS}$ mass given in
Ref.~\cite{PDG} is obtained from the pole mass at one loop.}  using
next-to-leading-order evolution equations (see the Appendix)
\cite{VLR}.\footnote{When the $b$ mass appears in the kinematics or
the perturbative heavy-quark distribution function [Eq.~(\ref{Q})],
the CTEQ mass (5.0 GeV) should be used.}  The value of the running $b$
mass at various values of $\mu_{UV}$ are listed in Table
\ref{tab_two}.  Using $\overline m_b(m_H)$, rather than $\overline
m_b(M_b)$ or $M_b$, to evaluate the Yukawa coupling decreases the
cross section by about $50\%$.  We use the standard-model Yukawa
coupling, with no enhancement factor, throughout.  Although there is a
lower bound of about 70 GeV on the mass of supersymmetric Higgs bosons
\cite{OPAL}, there is no strict lower bound on the mass of Higgs
bosons in a general two-Higgs-doublet model \cite{OPAL,KR}, so we
include the results for a few smaller masses.

We show the percentage change in the cross section from the corrections of 
order $1/\ln (m_H/m_b)$, $1/\ln^2 (m_H/m_b)$, and $\alpha_s$, as a function
of the Higgs-boson mass, in Figs.~5 (Tevatron) and 6 (LHC).
We find that the $1/\ln (m_H/m_b)$ correction is significant and negative
% at the Tevatron (LHC), ranging from $-86\%$ ($-79\%$) for 
% $m_H = 50$ GeV to $-35\%$ ($-36\%$) for $m_H = 1000$ GeV.
at the Tevatron (LHC), ranging from $-94\%$ ($-86\%$) for 
$m_H = 40$ GeV to $-35\%$ ($-36\%$) for $m_H = 1000$ GeV.
The size of this correction is a measure of the validity
of our calculation; as it approaches approximately 
$-100\%$, it is no longer justified to
regard $b\bar b \to H$ as the leading-order subprocess.  Our calculation 
is increasingly justified as one increases the Higgs mass, but the 
$1/\ln (m_H/m_b)$ correction is significant even for $m_H = 1000$ GeV.

\begin{figure}[tb]
\begin{center}
\epsfxsize= 3in  %actual
\leavevmode
\epsfbox{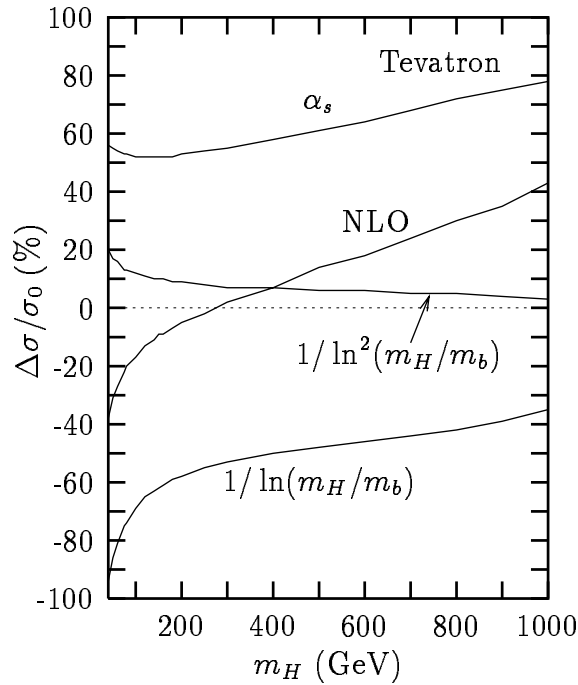}
\end{center}
\caption{Percentage change in the cross section for Higgs-boson production
in association with bottom quarks from the corrections of order
$1/\ln (m_H/m_b)$, $1/\ln^2 (m_H/m_b)$, and $\alpha_s$, as a function of the
Higgs-boson mass, at the Tevatron.  The next-to-leading-order (NLO) cross
section is the sum of the leading-order cross section and the corrections of
order $1/\ln (m_H/m_b)$ and $\alpha_s$.}
\end{figure}

\begin{figure}[tb]
\begin{center}
\epsfxsize= 3in  %actual
\leavevmode
\epsfbox{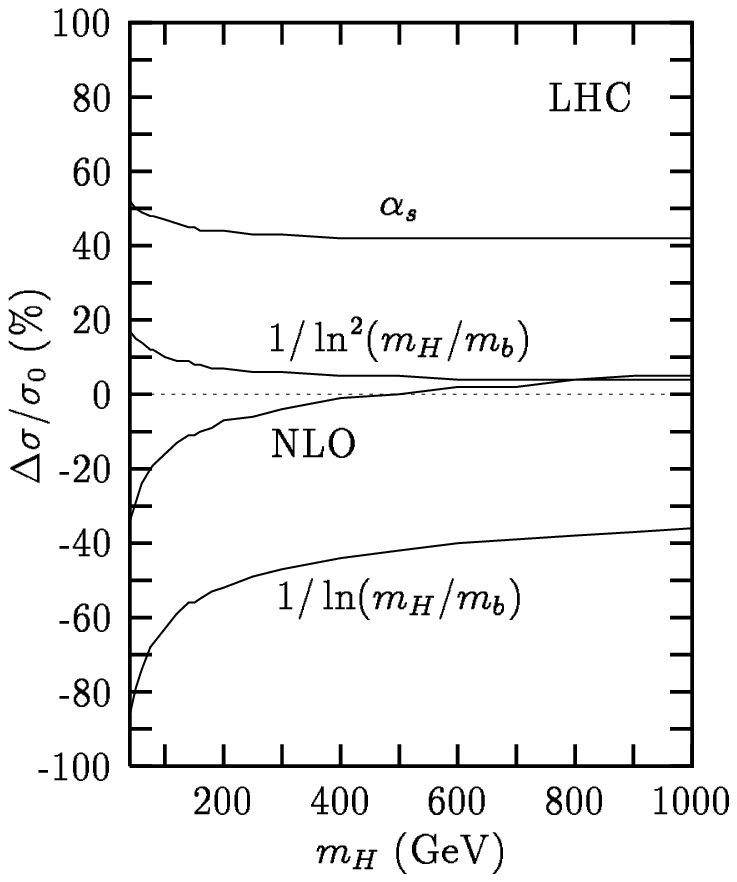}
\end{center}
\caption{Same as Fig.~5, but at the LHC.}
\end{figure}

The $\alpha_s$ correction is also significant, and happens to be positive,
such that it largely cancels the $1/\ln (m_H/m_b)$ correction.  The 
% $\alpha_s$ correction ranges from $+55\%$ for $m_H = 50$ GeV to $+78\%$
$\alpha_s$ correction ranges from $+56\%$ for $m_H = 40$ GeV to $+78\%$
for $m_H = 1000$ GeV at the Tevatron.  The correction increases as the 
Higgs-boson mass approaches the machine energy due to the presence of a large 
Sudakov logarithm \cite{S}.\footnote{We do not attempt
to sum the Sudakov logarithm.}  Such an effect is not
present at the much higher-energy LHC, where the correction ranges from
% $+50\%$ for $m_H = 50$ GeV to $+42\%$ for $m_H = 1000$ GeV.
$+52\%$ for $m_H = 40$ GeV to $+42\%$ for $m_H = 1000$ GeV.  

% The $1/\ln^2 (m_H/m_b)$ correction is modest, ranging from $+17\%$ ($+15\%$) 
% for $m_H = 50$ GeV to $+3\%$ ($+4\%$) 
The $1/\ln^2 (m_H/m_b)$ correction is modest, ranging from $+20\%$
($+17\%$) for $m_H = 40$ GeV to $+3\%$ ($+4\%$) for $m_H = 1000$ GeV
at the Tevatron (LHC). This supports our counting of the order of the
various corrections.  Since we have not calculated the other
next-to-next-to-leading-order corrections, of order $\alpha_s^2$ and
$\alpha_s \times 1/\ln (m_H/m_b)$, we do not include the $1/\ln^2
(m_H/m_b)$ correction in our final result, given in Table
\ref{tab_one}.

The largest uncertainty in the cross section comes from varying the
factorization scale, $\mu$.  We show in Figs.~7 (Tevatron) and 8 (LHC)
the percentage change in the cross section from its central value
($\mu = m_H$) due to varying $\mu$ between $m_H/2$ and $2m_H$, while
keeping $\mu_{UV} = m_H$ fixed.  The scale variation is generally
larger at next-to-leading order than it is at leading
order,\footnote{In these figures, the leading-order cross section is
calculated with leading-order parton distribution functions
\cite{CTEQ}.} which indicates that the leading-order scale variation
is not a reliable estimate of the uncertainty.  The
next-to-leading-order uncertainty ranges from about $\pm 30\%$ for
$m_H = 100$ GeV to about $\pm 8.5\%$ ($\pm 13\%$) for $m_H = 1000$ GeV
at the Tevatron (LHC).

\begin{figure}
\begin{center}
\epsfxsize= 3in  %actual
\leavevmode
\epsfbox{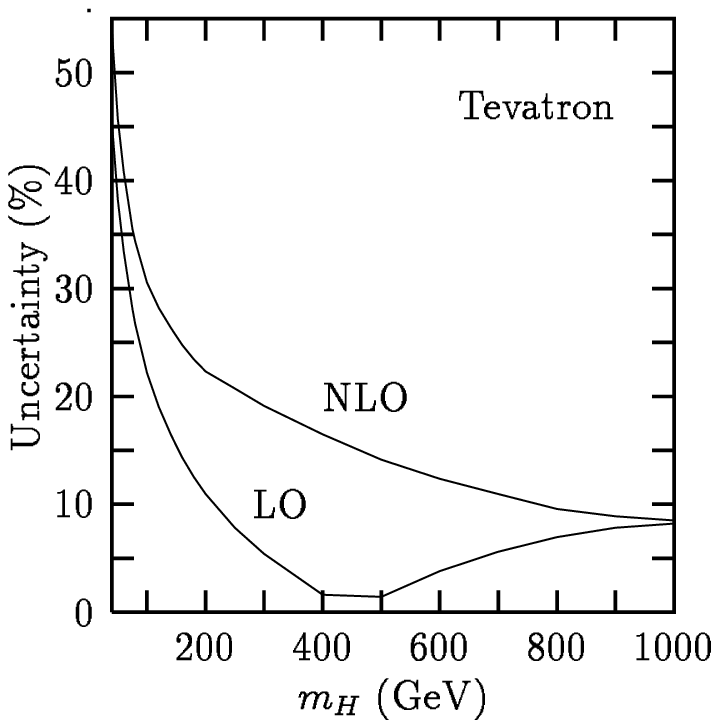}
\end{center}
\caption{Uncertainty in the cross section for Higgs-boson production in
association with bottom quarks at the Tevatron, obtained by 
varying the factorization scale $\mu$ about its central value, $\mu=m_H$,
from $m_H/2$ to $2m_H$.}
\end{figure}

\begin{figure}
\begin{center}
\epsfxsize= 3in  %actual
\leavevmode
\epsfbox{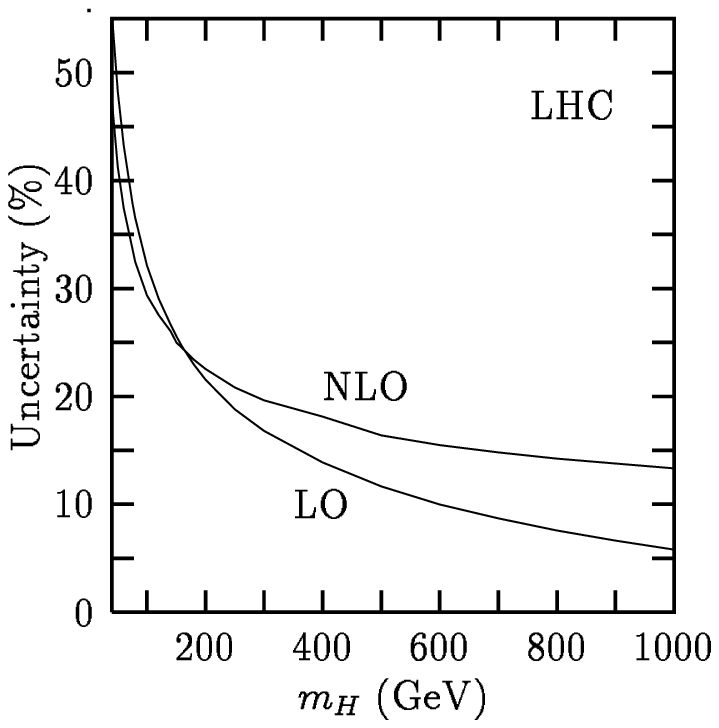}
\end{center}
\caption{Same as Fig.~7, but at the LHC.}
\end{figure}

There is a much smaller uncertainty in the cross section from varying
the renormalization scale, $\mu_{UV}$, between $m_H/2$ and $2m_H$.
The next-to-leading-order uncertainty ranges from about $\pm 2.5\%$
for $m_H = 100$ GeV to about $\pm 6\%$ ($\pm 4\%$) for $m_H = 1000$
GeV at the Tevatron (LHC).  This is significantly less than the
leading-order uncertainty of about $\pm 11\%$ at both machines.  There
is also an uncertainty in the cross section of about $\pm 8\%$ from
the uncertainty in the $b$-quark mass.

Since the $b$-quark distribution function arises from the gluon
distribution function, an additional source
of uncertainty is from the gluon-gluon luminosity, which depends on
$\tau \approx m_H^2/S$ \cite{CTEQG}.  We use the uncertainty advocated
in Ref.~\cite{CTEQG}: $\pm 10\%$ ($\sqrt \tau < 0.1$); $\pm 20\%$
($0.1 < \sqrt \tau < 0.2$); $\pm 30\%$ ($0.2 < \sqrt \tau < 0.3$);
$\pm 60\%$ ($0.3 < \sqrt \tau < 0.4$).  We combine all four sources of
uncertainty discussed above in quadrature, and report the uncertainty
in the next-to-leading-order cross sections in Table \ref{tab_one}.

Our calculation might also be applied to Higgs-boson production in
association with top quarks \cite{DW,K,GHPTW}. However, it is only valid
for $m_H \gg m_t$, where $t\bar t \to H$ can be regarded as leading
order and $gt \to Ht$ can be regarded as a small correction of order
$1/\ln(m_H/m_t)$.  For $m_H \sim m_t$, one must regard $gg \to t\bar
tH$ as the leading-order subprocess (along with $q\bar q \to t\bar t H$).
The $\alpha_s$ correction to these subprocesses has not yet been
calculated.\footnote{However, it has been
calculated in the opposite limit to the one we are considering, namely
$m_H \ll m_t$ (with $m_H,m_t \ll \sqrt {\hat s}$) \cite{DR}.}

In summary, we have calculated the cross section for Higgs-boson production 
in association with bottom quarks at next-to-leading-order in    
$1/\ln (m_H/m_b)$ and $\alpha_s$, and at next-to-next-to-leading order in 
$1/\ln (m_H/m_b)$.  The most important effect of the next-to-leading-order
corrections is taken into account by evaluating the bottom-quark Yukawa 
coupling using the $\overline{\rm MS}$ mass evaluated at the Higgs mass.
The $1/\ln (m_H/m_b)$ and $\alpha_s$ corrections are both large, but have the
opposite sign, such that the total next-to-leading-order correction is 
relatively modest. 

\begin{table}[tbp]
\begin{center}
\caption{Leading-order and next-to-leading-order cross sections (pb)
for Higgs-boson production in association with bottom quarks at the 
Tevatron ($\sqrt S =$ 1.8 and 2 TeV $p\bar p$) and the LHC ($\sqrt S =$ 
14 TeV $pp$).    The next-to-leading-order (NLO) cross section is the sum
of the leading-order cross section and the corrections of order
$1/\ln (m_H/m_b)$ and $\alpha_s$.
All calculations are performed in the $\overline{\rm MS}$
scheme using CTEQ4M parton distribution functions with $\mu=\mu_{UV}=m_H$.
The method to obtain the uncertainty in the next-to-leading-order 
cross section is described in the text. \label{tab_one}} \medskip
\hspace*{-0.5in}
\begin{tabular}{c|rll|rll|rll} \hline \hline
\multicolumn{1}{c}{}&\multicolumn{3}{c}{$\sqrt S = 1.8$~TeV $p\bar p$}&
\multicolumn{3}{c}{$\sqrt S = 2$~TeV $p\bar p$}&
\multicolumn{3}{c}{$\sqrt S = 14$~TeV $pp$} \\ 
\multicolumn{1}{c}{$m_H$ (GeV)} & \multicolumn{1}{c}{LO} &
\multicolumn{1}{c}{NLO} &
\multicolumn{1}{c}{(pb)} & \multicolumn{1}{c}{LO} &
\multicolumn{1}{c}{NLO} & \multicolumn{1}{c}{(pb)}
&\multicolumn{1}{c}{LO} & \multicolumn{1}{c}{NLO}&
\multicolumn{1}{c}{(pb)} \\ \hline
%$m_H$ (GeV)&\multicolumn{2}{c}{$\sqrt S = 1.8$~TeV $p\bar p$ (pb)}& \multicolumn{2}{c}{$\sqrt S = 2$~TeV $p\bar p$ (pb)}& \multicolumn{2}{c}{$\sqrt S = 14$~TeV $pp$ (pb)} \\
%& \multicolumn{1}{c}{LO} & \multicolumn{1}{c}{NLO} & \multicolumn{1}{c}{LO} & \multicolumn{1}{c}{NLO} & \multicolumn{1}{c}{LO} & \multicolumn{1}{c}{NLO} \\ \hline
40 & 12.30 &  $7.45\pm 4.17$ & $\times 10^{-1}$ & 15.10 & $9.36\pm 5.19$ & $\times 10^{-1}$ & 3.16 & $2.10\pm 1.03$ & $\times 10^{1}$\\
% 50 & 5.49 &  $3.75\pm 1.80$ & $\times 10^{-1}$ & 6.81 & $4.69\pm 2.23$ & $\times 10^{-1}$ & 1.70 & $1.21\pm 0.53$ & $\times 10^{1}$\\
60 & 2.71 &  $1.99\pm 0.85$ & $\times 10^{-1}$ & 3.41 & $2.50\pm 1.07$ & $\times 10^{-1}$ & 9.94 & $7.52\pm 2.98$ & $\times 10^{0}$\\
% 75 & 10.8 &  $8.45\pm 3.22$ & $\times 10^{-2}$ & 1.40 & $1.09\pm 0.41$ & $\times 10^{-1}$ & 5.05 & $4.05\pm 1.46$ & $\times 10^{0}$\\
80 & 8.23 &  $6.52\pm 2.41$ & $\times 10^{-2}$ & 10.70 & $8.52\pm 3.13$ & $\times 10^{-2}$ & 4.12 & $3.33\pm 1.16$ & $\times 10^{0}$\\
100 & 3.05 & $2.56\pm 0.86$ & $\times 10^{-2}$ & 4.05 & $3.38\pm 1.12$ & $\times 10^{-2}$ & 2.01 & $1.69\pm 0.54$ & $\times 10^{0}$\\
120 & 1.28 & $1.11\pm 0.35$ & $\times 10^{-2}$ & 1.74 & $1.51\pm 0.47$ & $\times 10^{-2}$ & 11.00 & $9.55\pm 2.92$ & $\times 10^{-1}$\\
140 & 5.90 & $5.28\pm 1.57$ & $\times 10^{-3}$ & 8.22 & $7.33\pm 2.17$ & $\times 10^{-3}$ & 6.47 & $5.76\pm 1.68$ & $\times 10^{-1}$\\
% 150 & 4.12 & $3.75\pm 1.09$ & $\times 10^{-3}$ & 5.80 & $5.26\pm 1.52$ & $\times 10^{-3}$ & 5.04 & $4.48\pm 1.27$ & $\times 10^{-1}$\\
160 & 2.91 & $2.67\pm 0.75$ & $\times 10^{-3}$ & 4.15 & $3.79\pm 1.07$ & $\times 10^{-3}$ & 4.02 & $3.61\pm 1.01$ & $\times 10^{-1}$\\
180 & 1.51 & $1.42\pm 0.39$ & $\times 10^{-3}$ & 2.21 & $2.06\pm 0.56$ & $\times 10^{-3}$ & 2.63 & $2.39\pm 0.65$ & $\times 10^{-1}$ \\
200 & 8.21 & $7.83\pm 2.49$ & $\times 10^{-4}$ & 1.23 & $1.17\pm 0.31$ & $\times 10^{-3}$ & 1.78 & $1.65\pm 0.43$ & $\times 10^{-1}$ \\
250 & 2.04 & $2.03\pm 0.61$ & $\times 10^{-4}$ & 3.24 & $3.19\pm 0.97$ & $\times 10^{-4}$ & 7.66 & $7.23\pm 1.79$ & $\times 10^{-2}$\\
300 & 5.84 & $6.01\pm 1.75$ & $\times 10^{-5}$ & 0.98 & $1.00\pm 0.29$ & $\times 10^{-4}$ & 3.74 & $3.58\pm 0.85$ & $\times 10^{-2}$\\
400 & 6.10 & $6.67\pm 2.36$ & $\times 10^{-6}$ & 1.17 & $1.25\pm 0.35$ & $\times 10^{-5}$ & 1.15 & $1.14\pm 0.26$ & $\times 10^{-2}$\\
500 & 7.70 & $8.86\pm 3.05$ & $\times 10^{-7}$ & 1.70 & $1.93\pm 0.67$ & $\times 10^{-6}$ & 4.34 & $4.34\pm 0.92$ & $\times 10^{-3}$\\
600 & 1.07 & $1.29\pm 0.80$ & $\times 10^{-7}$ & 2.75 & $3.25\pm 1.10$ & $\times 10^{-7}$ & 1.90 & $1.93\pm 0.40$ & $\times 10^{-3}$ \\
700 & 1.54 & $1.97\pm 1.21$ & $\times 10^{-8}$ & 4.72 & $5.83\pm 3.60$ & $\times 10^{-8}$ & 9.21 & $9.44\pm 1.89$ & $\times 10^{-4}$\\
800 & 2.28 & $3.08\pm 1.90$ & $\times 10^{-9}$ & 0.83 & $1.08\pm 0.66$ & $\times 10^{-8}$ & 4.79 & $4.97\pm 0.98$ & $\times 10^{-4}$\\
900 & 3.42 & $4.87\pm 3.00$ & $\times 10^{-10}$ & 1.49 & $2.01\pm 1.24$ & $\times 10^{-9}$ & 2.64 & $2.77\pm 0.54$ & $\times 10^{-4}$\\
1000 & 5.18 &$7.82\pm 4.81$ & $\times 10^{-11}$ & 2.70 & $3.85\pm 2.37$ & $\times 10^{-10}$ & 1.52 & $1.60\pm 0.31$ & $\times 10^{-4}$\\
\hline \hline
\end{tabular}
\end{center}
\end{table}

\begin{table}[tbp]
\begin{center}
\caption{The two-loop running $b$ mass at various values of the 
renormalization scale $\mu_{UV}$, given the initial value
$\overline m_b(M_b)=4.25$ GeV. \label{tab_two}} \medskip
\begin{tabular*}{\textwidth}{c@{\extracolsep\fill}ccc} \hline \hline
&$\mu_{UV}$ (GeV) & $\overline m_b(\mu_{UV})$ (GeV) &\\ \hline
& 40 & 3.23 & \\ &
%50 & 3.17 & \\ &
60 & 3.11 & \\ &
%75 & 3.05 & \\ &
80 & 3.04 & \\ &
100 & 2.98 & \\ & 
120 & 2.94 & \\ &
140 & 2.90 & \\ &
%150 & 2.89 & \\ & 
160 & 2.87 & \\ &
180 & 2.85 & \\ &
200 & 2.82 & \\ & 
250 & 2.77 & \\ & 
300 & 2.73 & \\ & 
400 & 2.68 & \\ & 
500 & 2.63 & \\ & 
600 & 2.60 & \\ & 
700 & 2.57 & \\ &
800 & 2.55 & \\ & 
900 & 2.53 & \\ & 
1000 & 2.51 & \\ 
\hline \hline
\end{tabular*}
\end{center}
\end{table}

\section*{Acknowledgments}

\indent\indent We are grateful for conversations and correspondence with
M.~Krawczyk, M.~Oreglia, M.~Seymour, and T.~Sjostrand.
S.~W.~thanks the Enrico Fermi Institute for support. 
This work was supported in part by the U.~S.~Department of Energy,
High Energy Physics Division, under Contract W-31-109-Eng-38 and Grant Nos.
DE-FG013-93ER40757 and DE-FG02-91ER40677.

\section*{Appendix}

The next-to-leading-order running of the heavy-quark $\overline{\rm
MS}$ mass is described by the equation \cite{VLR}
\begin{equation}
\overline m(\mu) = \overline m(\mu_0)
\left(\frac{\alpha_s(\mu)}{\alpha_s(\mu_0)}\right)^{\frac{\gamma_0}
{\beta_0}}
\frac{[1+A_1\alpha_s(\mu)/\pi]}
     {[1+A_1\alpha_s(\mu_0)/\pi]} \;,
\label{BRNG2}
\end{equation}
where 
\begin{equation}
A_1 = -\frac{\beta_1\gamma_0}{\beta_0^2} + \frac{\gamma_1}{\beta_0} \;,
\end{equation}
\begin{eqnarray*}
\beta_0 &=& \frac{1}{4}\left(11-\frac{2}{3}N_f\right) \;\;\;\;\;\;,\;
\gamma_0 = 1 \;, \\
\beta_1 &=& \frac{1}{16}\left(102-\frac{38}{3}N_f\right) \;,\;
\gamma_1 = \frac{1}{16}\left(\frac{202}{3}-\frac{20}{9}N_f\right) \;,
\end{eqnarray*}
where $N_f$ is the number of quark flavors of mass less than $\mu$ 
($\mu > \mu_0$).  For $\mu < m_t$, $N_f = 5$; for $\mu > m_t$, $N_f =6$,
with the running mass continuous at $\mu=m_t$.

The relation between the pole mass and the $\overline{\rm MS}$ mass at
next-to-leading order is
\begin{equation}
M_Q = \overline m_Q(M_Q)\left(1+C_F\frac{\alpha_s}{\pi}\right)\;.
\end{equation}

The next-to-leading-order evolution of the strong coupling is supplied as
a subroutine by CTEQ \cite{CTEQ}.  An analytic expression can be found
in Ref.~\cite{PDG}.

\end{document}